\documentclass[12pt,preprint]{aastex}

\begin{document}

\title{Simultaneous Triggered Collapse of the Presolar Dense Cloud Core 
and Injection of Short-Lived Radioisotopes by a Supernova Shock Wave}

\author{Alan P.~Boss\altaffilmark{1}, Sergei I. Ipatov\altaffilmark{1}, 
Sandra A. Keiser\altaffilmark{1}, Elizabeth A. Myhill\altaffilmark{1,2}, 
and Harri A. T. Vanhala\altaffilmark{1,3}}
\altaffiltext{1}{Department of Terrestrial Magnetism, Carnegie Institution of
Washington, 5241 Broad Branch Road, NW, Washington, DC 20015-1305;
boss@dtm.ciw.edu, siipatov@hotmail.com, keiser@dtm.ciw.edu.}
\altaffiltext{2}{Marymount University, Arlington, VA 22207;
elizabeth.myhill@marymount.edu.}
\altaffiltext{3}{Universities Space Research Association, Columbia, MD 
21044; hvanhala@usra.edu.}

\begin{abstract}

 Cosmochemical evidence for the existence of short-lived radioisotopes
(SLRI) such as $^{26}$Al and $^{60}$Fe at the time of the formation of 
primitive meteorites requires that these isotopes were synthesized in a 
massive star and then incorporated into chondrites within $\sim 10^6$ yr.
A supernova shock wave has long been hypothesized to have transported 
the SLRI to the presolar dense cloud core, triggered 
cloud collapse, and injected the isotopes. Previous numerical
calculations have shown that this scenario is plausible when the
shock wave and dense cloud core are assumed to be isothermal at
$\sim 10$ K, but not when compressional heating to $\sim 1000$ K is
assumed. We show here for the first time that when calculated with
the FLASH2.5 adaptive mesh refinement (AMR) hydrodynamics code, a 20 km/sec
shock wave can indeed trigger the collapse of a 1 $M_\odot$ cloud
while simultaneously injecting shock wave isotopes into the collapsing
cloud, provided that cooling by molecular species such as H$_2$O,
CO$_2$, and H$_2$ is included. These calculations imply that the
supernova trigger hypothesis is the most likely mechanism for delivering 
the SLRI present during the formation of the solar system.

\end{abstract}


\section{Introduction}

 The discovery of evidence for live $^{26}$Al (half-life of 
$0.7 \times 10^6$ yr) at the time of the 
formation of refractory inclusions in the Allende meteorite
(Lee et al. 1976) led quickly to the suggestion that the
$^{26}$Al was synthesized in a supernova, then transported
by the supernova's shock wave to a dense molecular cloud, 
where the shock wave triggered the collapse of a cloud core
and injected the $^{26}$Al into the collapsing presolar cloud 
(Cameron and Truran 1977). Detailed numerical calculations
showed that such simultaneous triggering and injection was possible,
provided that the shock wave had slowed to speeds of $\sim$ 10 to
$\sim$ 40 km/sec (Boss 1995; Foster \& Boss 1996, 1997) and that
the shock wave and cloud were effectively isothermal at $\sim$ 10 K.
This isothermal assumption is appropriate for the radiative phase
of a supernova shock (Chevalier 1974), which occurs when the shock has
travelled $\sim$ 10 pc and swept up a cool shell of gas and dust.
Recent adaptive mesh refinement (AMR) studies (Nakamura et al. 2006; 
Melioli et al. 2006) have shown that shock-triggered star formation is 
likely to occur 
when the supernova shock has evolved into a radiative shock, i.e., the
shock wave is able to cool so rapidly by radiation that the shock 
wave is effectively at the same temperature as the ambient medium, 
confirming the results of Boss (1995) and Foster \& Boss (1996, 1997).

 Vanhala \& Cameron (1998; hereafter VC) used a smoothed 
particle hydrodynamics (SPH) code and detailed equations of state
to confirm the previous results for isothermal shocks and clouds.
However, when VC allowed their models to become nonisothermal
by including compressional heating and molecular, atomic, and
dust cooling, they found that they could not achieve simultaneous
triggering of cloud collapse and injection of shock wave
material, raising serious doubts about the supernova trigger
hypothesis.

 Alternative explanations for the origin of the SLRI in 
chondritic meteorites have also arisen. Shu et al. (1997)
suggested that the SLRI were produced {\it in situ} by spallation 
reactions involving energetic particles emanating from protosolar 
flares striking dust grains near the X-wind point, from whence
the grains would be lofted upward by the X-wind and transported
outward to the solar nebula. However, the SLRI $^{60}$Fe 
cannot be produced by spallation, and requires a stellar
nucleosynthetic source (Tachibana \& Huss 2003) such as a
supernova. The fact that the results of $^{26}$Al dating agree 
well with the independent Pb-Pb dating system (Connelly et al. 2008)
implies that the $^{26}$Al was spatially homogeneous in the 
solar nebula, which is inconsistent with the bulk of its production
by spallation reactions at the X-wind point. Regardless of
whether spallation reactions contributed to the SLRI inventory,
then, a supernova source for the $^{26}$Al and $^{60}$Fe appears
to be necessary. 
 
 Given a supernova origin for the $^{26}$Al and $^{60}$Fe,
it has also been suggested that a nearby ($\sim$ 0.1 pc) supernova 
may have injected these SLRI into the already-formed solar nebula 
(Ouellette, Desch \& Hester 2007). While such hot shock front 
gas is unable to penetrate into the much denser disk gas 
(Chevalier 2000), Ouellette et al. (2007) suggested that SLRI 
residing in dust grains micron-sized and larger shot through the stalled 
shock-wave gas and into the disk. However, supernova dust grains
are essentially all smaller than 0.1 micron, and are sputtered to even 
smaller sizes in the shock (Bianchi \& Schneider 2007). Gounelle \& 
Meibom (2008) noted that disks formed within $\sim$ 0.3 pc of a massive 
star would be photoevaporated away prior to its supernova explosion.
Furthermore, Krot et al. (2008) argued that injection into a late-phase 
solar nebula would have led to the injection of stable oxygen isotopes 
as well as SLRI into the disk, leading to an oxygen isotope 
distribution that would be inconsistent with the observed values and
their explanation by mass-independent fractionation. Krot et al. 
(2008) and Thrane et al. (2008) argued that injection must
have occurred instead into the presolar cloud, so that the sun and
the solar nebula shared a common reservoir of oxygen isotopes.

 Here we return to the question of whether nonisothermal shock
fronts are fatal for the supernova triggering hypothesis, by using
the FLASH2.5 AMR code and an improved cooling law to reinvestigate 
this basic question of solar system origin.
            
\section{Numerical Methods and Initial Conditions}

 FLASH employs a block-structured adaptive grid approach.
Advection is handled by the piecewise parabolic method (PPM), 
featuring a Riemann solver at cell boundaries that handles shock 
fronts exceptionally well. We downloaded and ran the FLASH2.5 
AMR code (flash.uchicago.edu/website/home) on several of the
FLASH-supplied test problems relevant to shock-triggered star
formation, namely the Sod shock tube problem and the gravitational
collapse of a pressureless cloud.
We then used FLASH2.5 to reproduce the standard case of
triggered collapse of Foster \& Boss (1996), and verified
that FLASH2.5 was able to produce simultaneous triggered collapse
and injection of shock wave material when the shocked cloud was
forced to remain isothermal at 10 K. The details of these test 
cases will be presented in a future paper.

 In the present models, we used the two dimensional, cylindrical 
coordinate ($R, Z$) version of FLASH2.5, with axisymmetry about 
the rotational axis ($\hat z$). Multipole gravity was used, 
including Legendre polynomials up to $l = 10$. The equation
of state routines were taken to be those for a simple
perfect gas with a mean molecular weight of $\mu = 2.3$.

 As in Foster \& Boss (1996), the target dense cloud core is 
a stable Bonnor-Ebert (BE) sphere with a mass of 1 $M_\odot$, a 
radius of 0.058 pc, a temperature of $T = 10$ K, and a maximum density of 
$6.2 \times 10^{-19}$ g cm$^{-3}$ at rest near the top of
the cylindrical grid. The BE sphere is embedded in an intercloud
medium with a density of $3.6 \times 10^{-22}$ g cm$^{-3}$ 
and a temperature of 10 K. The shock wave begins at the top of
the grid and propagates downward at 20 km/sec toward the
BE sphere. The shock wave has a thickness of 0.003 pc with a 
uniform density of $3.6 \times 10^{-20}$ g cm$^{-3}$ and a
temperature of 1000 K. The shock wave is followed by a
wind with a density of $3.6 \times 10^{-22}$ g cm$^{-3}$ 
and temperature of 1000 K also moving downward at 20 km/sec. 
The shock wave material is represented by a color field, initially 
defined to be equal to 1 inside the shock wave and 0 elsewhere,
which allows the shock wave material to be tracked in time 
(Foster \& Boss 1997). The SLRI are assumed to be contained in 
dust grains of sub-micron size (Bianchi \& Schneider 2007), 
small enough for the grains to remain coupled to the gas. 
These initial conditions are identical to those in the standard
case of Foster \& Boss (1996), with the exception of the
temperatures of 1000 K in the shock and wind, and the
nonzero velocity of the wind.

 The cylindrical grid is 0.197 pc long in $Z$ and 0.063 pc
wide in $R$. We set the number of blocks in $R$ to be 5 and
in $Z$ to be 15, leading to approximately uniform spacing in $R$
and $Z$, with each block consisting of $8 \times 8$ grid
points, equivalent to an initial grid of $40 \times 120$.
With up to five levels of refinement allowed, FLASH is
then able to follow small-scale structures with the
effective resolution of a grid 16 times finer in scale, or
effectively $640 \times 1920$, somewhat higher than the highest 
resolution of $480 \times 1440$ used by Vanhala \& Boss (2000), 
but less than the resolution of $960 \times 2880$ used by 
Vanhala \& Boss (2002).

\section{Results}

  In the absence of cooling, FLASH produces an adiabatic evolution 
with an effective $\gamma = 5/3$. Model NC was run without cooling, 
but with the temperature constrained to be less than 1000 K. 
The shock wave was unable to compress any part of the target 
cloud to a density higher than about twice that of the initial
central density of the cloud. Much of the cloud was heated to 
temperatures of 100 K to 1000 K, preventing cloud collapse.
Instead, the cloud's remnants were swept up by the shock wave and
wind and transported off the grid. This result is quite similar to
that obtained by Foster \& Boss (1996) for the standard case 
when run with an adiabatic pressure law ($\gamma = 5/3$) rather
than an isothermal pressure law.
 
 We next ran three models with variations in the cooling
function, based on the results of Neufeld \& Kaufman (1993),
who calculated the radiative cooling caused by rotational
and vibrational transitions of optically thin, warm molecular gas 
composed of H$_2$O, CO, and H$_2$, finding H$_2$O to be the
dominant cooling agent. Their Figure 3 shows that over
the range of temperature from 100 K to 4000 K, the total
cooling rate coefficient $L$ can be approximated as
$L \approx L_0 \approx 10^{-24} (T/100)$ erg cm$^3$ s$^{-1}$.
The cooling rate $\Lambda = L \ n(H_2) \ n(m)$, where $n(H_2)$
is the number density of molecular hydrogen and $n(m)$ is
the number density of the molecular species under consideration.
Assuming that $n(H_2O)/n(H_2) \approx 8.8 \times 10^{-4}$,
we take $n(m)/n(H_2) \approx 10^{-3}$, leading to 
$\Lambda \approx 9 \times 10^{19} (T/100) \rho^2$ erg cm$^{-3}$ s$^{-1}$,
where $\rho$ is the gas density in g cm$^{-3}$.

 The figures show the results of model C, which used the Neufeld \& 
Kaufman (1993) cooling rate as well as the constraint that the 
temperatures remain at 1000 K or less. Figure 1 shows that Rayleigh-Taylor 
(R-T) fingers form immediately after the shock strikes the cloud
and penetrate downward farthest along the symmetry axis. Kelvin-Helmholtz 
(K-H) rolls form soon thereafter, as the shock front ablates material
off the sides of the cloud and transports it downstream.
The color field is injected into the target cloud by a 
combination of the R-T and K-H instabilities, though the
cloud material polluted by the K-H vortices tends to be
lost by subsequent ablation; the R-T fingers have
the best chance to inject shock wave material close to the
cloud's symmetry axis and hence into the region where
a protostar will soon form. The fact that both gas and dust from
the shock wave region are directly injected into the cloud
by R-T fingers ensures that SLRI carried by the shock wave 
will also be injected, as found in the previous isothermal 
models (Foster \& Boss 1997; Vanhala \& Boss 2000, 2002)

 Figure 2 shows that after 0.1 Myr, a region along
the symmetry axis has formed with a maximum density $\sim$ 1000
times that of the center of the initial target cloud. The temperature
contours show that the thermal energy generated by compressional 
heating at the shock-cloud interface is rapidly lost by the
molecular cooling. The maximum temperatures of 1000 K are
limited to thin shells at the shock-cloud interface; cooling
is so rapid in the denser regions just inside this interface
that the temperature falls to 10 K, the minimum temperature
allowed by the calculation. This rapid cooling makes the
evolution similar to that of isothermal models.

 Figures 3 and 4 show a close-up of the forming protostar
along the symmetry axis, after 0.16 Myr, when the maximum
density has reached $2 \times 10^{-12}$ g cm$^{-3}$. Lower
density regions continue to infall onto the growing central
protostar with velocities on the order of 1 km/sec, highly
supersonic compared to the sound speed for 10 K gas of
0.2 km/sec. Clearly the protostar has entered the dynamic
collapse phase, and will form a central protostellar
core surrounded by an infalling envelope. Figure 4 shows 
that while the protostar is polluted with shock wave material, 
the lower density gas that will soon accrete onto the protostar 
typically has an even higher density of shock wave material 
than that already in the protostar. Note that the
present models do not include rotation of the target cloud,
so a protoplanetary disk cannot form in these models.

 We estimate the injection efficiency of the shock wave material
into the forming protostar by calculating the amount of color
residing inside regions with density greater than $10^{-18}$ g cm$^{-3}$
at the time (0.16 Myr) shown in Figures 3 and 4. The fraction
of the incident color field that was injected into this infalling
region is $\sim$ 0.003 for model C. This low injection efficiency
is similar to that ($\sim$ 0.002) obtained by Vanhala \& Boss (2002) 
in their extremely high resolution, isothermal shock-cloud models,
for a comparable definition of injection efficiency.

 Such a low injection efficiency is in accord with a supernova 
as the source of the shock wave. Based on the estimates of 
Cameron et al. (1995), Foster \& Boss (1997) noted 
that the $^{26}$Al-containing gas and dust in a supernova shock 
wave would have to be diluted by a factor of $\sim 10^4$ in order
to explain the inferred initial abundance of $^{26}$Al in the
solar nebula, i.e., $10^{-4} M_\odot$ of shock-wave material
should be injected into a 1 $M_\odot$ presolar cloud. 
In the standard case, the mass of the shock wave 
that is incident on the target cloud is 0.016 $M_\odot$,
so the desired injection efficiency is $\sim 0.006$. However,
given that not all of the 1 $M_\odot$ target cloud will be
accreted by the protostar, the injection efficiency would
need to be half as large if only 0.5 $M_\odot$ is accreted,
dropping the desired injection efficiency to $\sim 0.003$,
the same as that obtained in model C. Many other factors
enter into the desired injection efficiency, such as radioactive
decay, nucleosynthetic yields (Rauscher et al. 2002),
and sweeping-up of intervening interstellar cloud material,
so these estimates should only be taken as being consistent
to order of magnitude. 

 Two other models with cooling were calculated, identical to model C 
except for having the Neufeld \& Kaufman (1993) cooling rate doubled 
(model 2C) or halved (model 0.5C), in order to test the sensitivity of
the results to the assumed cooling rate. Both models evolved
very similarly to model C, with the main difference being
that the amount of injected shock front material was about
1/3 higher in model 2C and about 1/10 lower in model 0.5C.
Evidently simultaneous triggering and injection is relatively
insensitive to changes in $\Lambda$ of this magnitude.

 Finally, it is worth considering why these results differ from
those of VC, who were unable to achieve simultaneous triggering
and injection in their three dimensional SPH models with cooling. 
The reasons appear to be three-fold. First, the FLASH AMR code with 
PPM is superb at representing the physics of shock waves striking
target clouds, such as the R-T and K-H instabilities that 
dominate these interactions. SPH has its strengths, but SPH is
relatively poor at resolving the R-T and K-H dynamical 
instabilities (e.g., Agertz et al. 2007). Second, the 
Neufeld \& Kaufman (1993) cooling rate estimate is a considerable
improvement over the cooling rates used in VC, which included
many species, but did not include the contribution due to H$_2$O,
which Neufeld \& Kaufman (1993) found to be dominant. Third, and
perhaps most importantly, low injection efficiencies would be hard 
to detect in the VC SPH calculations, which typically involved
only 5000 particles in the target cloud, so that the desired dilution
factor of $10^4$ for a supernova shock wave (Foster \& Boss 1997) 
would require that only half a particle be injected. 

\section{Conclusions}

 These models show that when cooling by molecular species is
included, a 20 km/sec shock wave is able to trigger the gravitational
collapse of an otherwise stable, solar-mass dense cloud core, as
well as to inject appropriate amounts of supernova shock wave material
into the collapsing cloud core. This injected material consists
of shock wave gas as well as dust grains small enough to remain
coupled to the gas, i.e., sub-micron-sized grains, which are
expected to characterize supernova shock waves (Bianchi \& 
Schneider 2007) and to carry the SLRI whose decay products have 
been found in refractory inclusions of chondritic meteorites
(Lee et al. 1976). Evidently a radiative-phase supernova shock 
wave is able to cool sufficiently rapidly to behave in much 
the same way as a shock wave that is assumed to remain isothermal 
with the target cloud (Boss 1995; Foster \& Boss 1996, 1997;
Vanhala \& Boss 2000, 2002). These models thus lend strong support 
to the hypothesis first advanced by Cameron \& Truran (1977) that
a supernova shock wave carrying SLRI may have triggered the 
formation of the solar system. 

\acknowledgements

 We thank Tomasz Plewa for valuable assistance with using FLASH2.5
and Roger Chevalier, James Stone, and the two referees for their comments.
This research was supported in part by NASA Origins of Solar Systems 
grant NNG05GI10G and NASA Planetary Geology and Geophysics grant 
NNX07AP46G, and is contributed in part to NASA Astrobiology Institute
grant NCC2-1056. The software used in this work was in part
developed by the DOE-supported ASC/Alliances Center for 
Astrophysical Thermonuclear Flashes at the University of Chicago.

\clearpage

\begin{figure}
\vspace{-1.0in}
\plotone{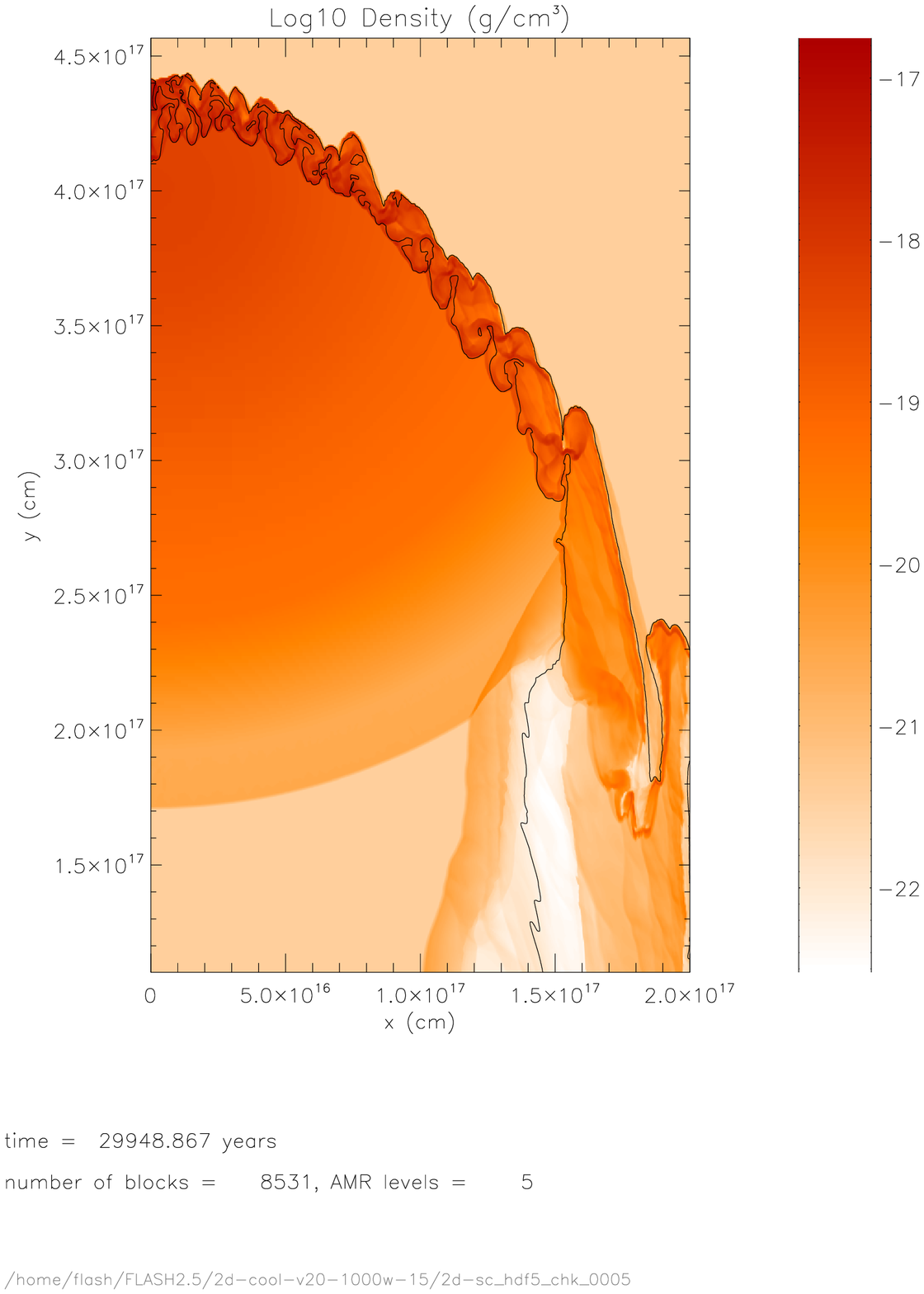}
\vspace{+0.25in}
\caption{Density for model C after 0.03 Myr of evolution. Contours 
show regions with color fields (representing SLRI) greater than 0.01.
Symmetry axis is along the left hand side of the plot. Shock
wave travels downward from top of box. R-T fingers and K-H
vortices form at the shock-cloud interface. $R$ axis is horizontal
and $Z$ axis is vertical.}
\end{figure}

\begin{figure}
\vspace{-1.0in}
\plotone{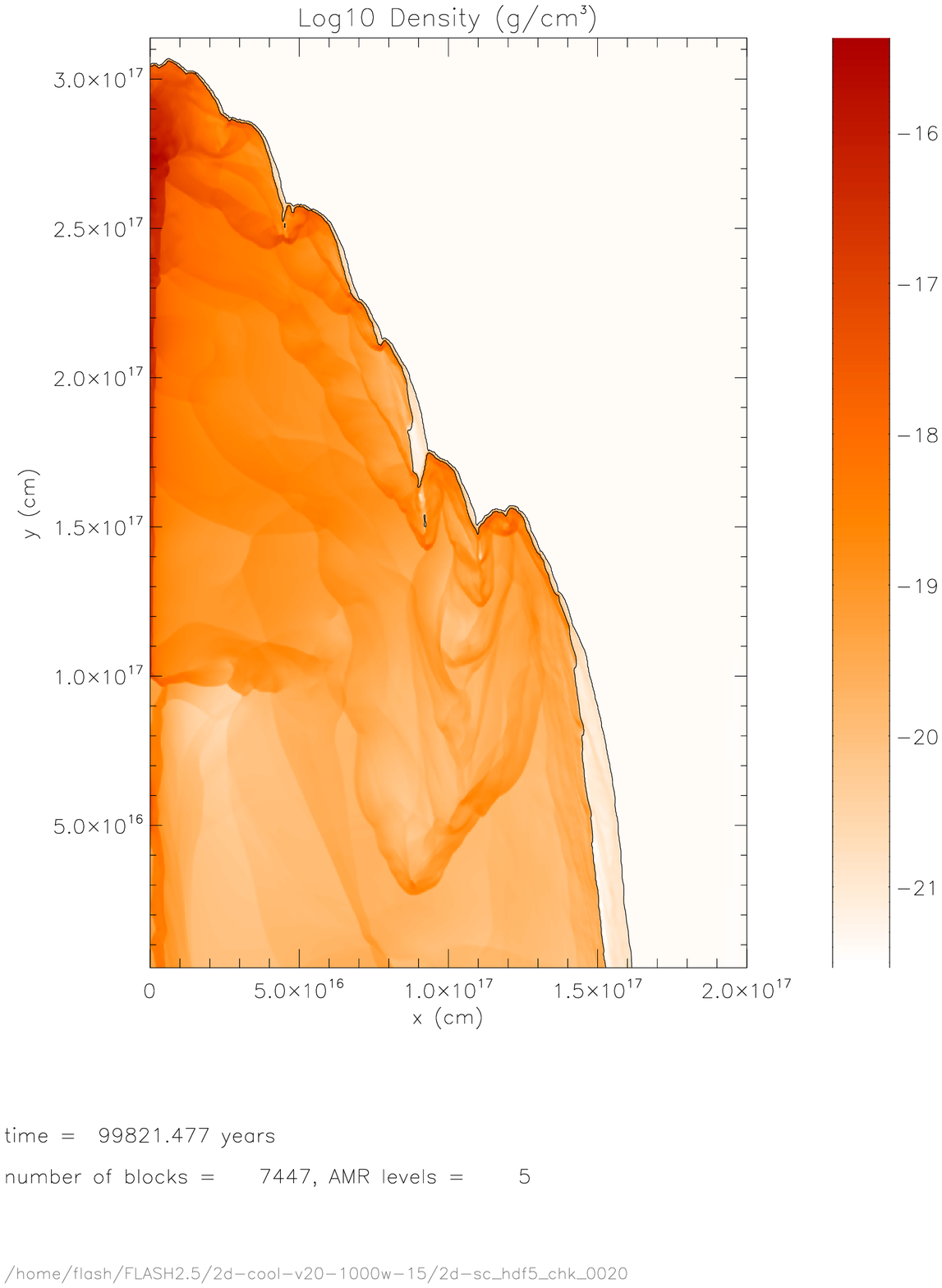}
\vspace{+0.25in}
\caption{Same as Figure 1 except after 0.1 Myr. Contours now
show regions with temperatures greater than 100 K, which only
occur at the shock-cloud interface as a result of the molecular
cooling. A high-density region has formed along the symmetry axis.}
\end{figure}

\begin{figure}
\vspace{-1.0in}
\plotone{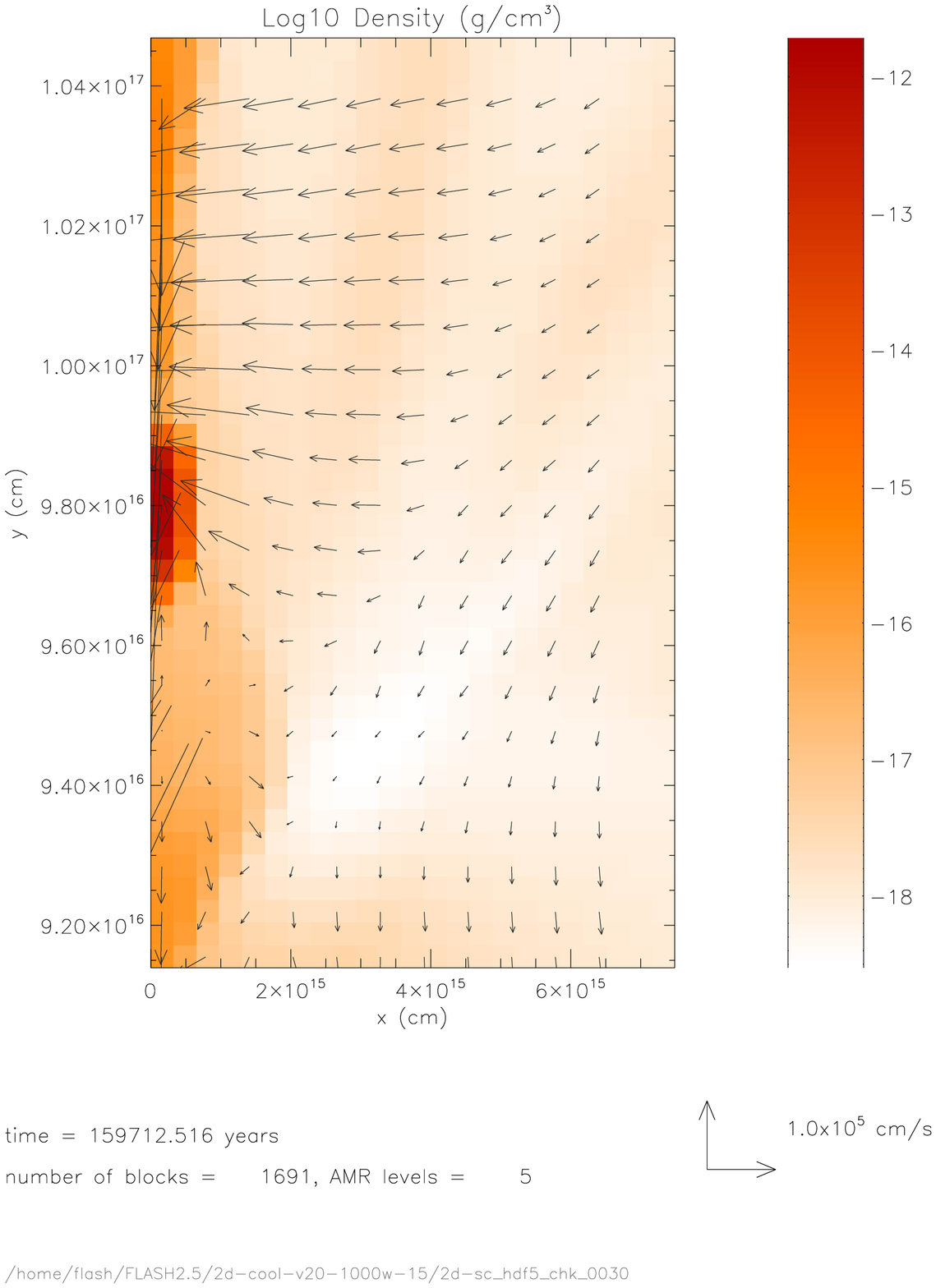}
\vspace{+0.25in}
\caption{Same as Figure 1 except after 0.16 Myr and limited to a small
region around the density maximum of $\sim 2 \times 10^{-12}$ g cm$^{-3}$.
Velocity contours are shown for every other AMR grid cell. Much 
of the cloud is infalling onto the growing protostar on the 
symmetry axis.}
\end{figure}

\begin{figure}
\vspace{-1.0in}
\plotone{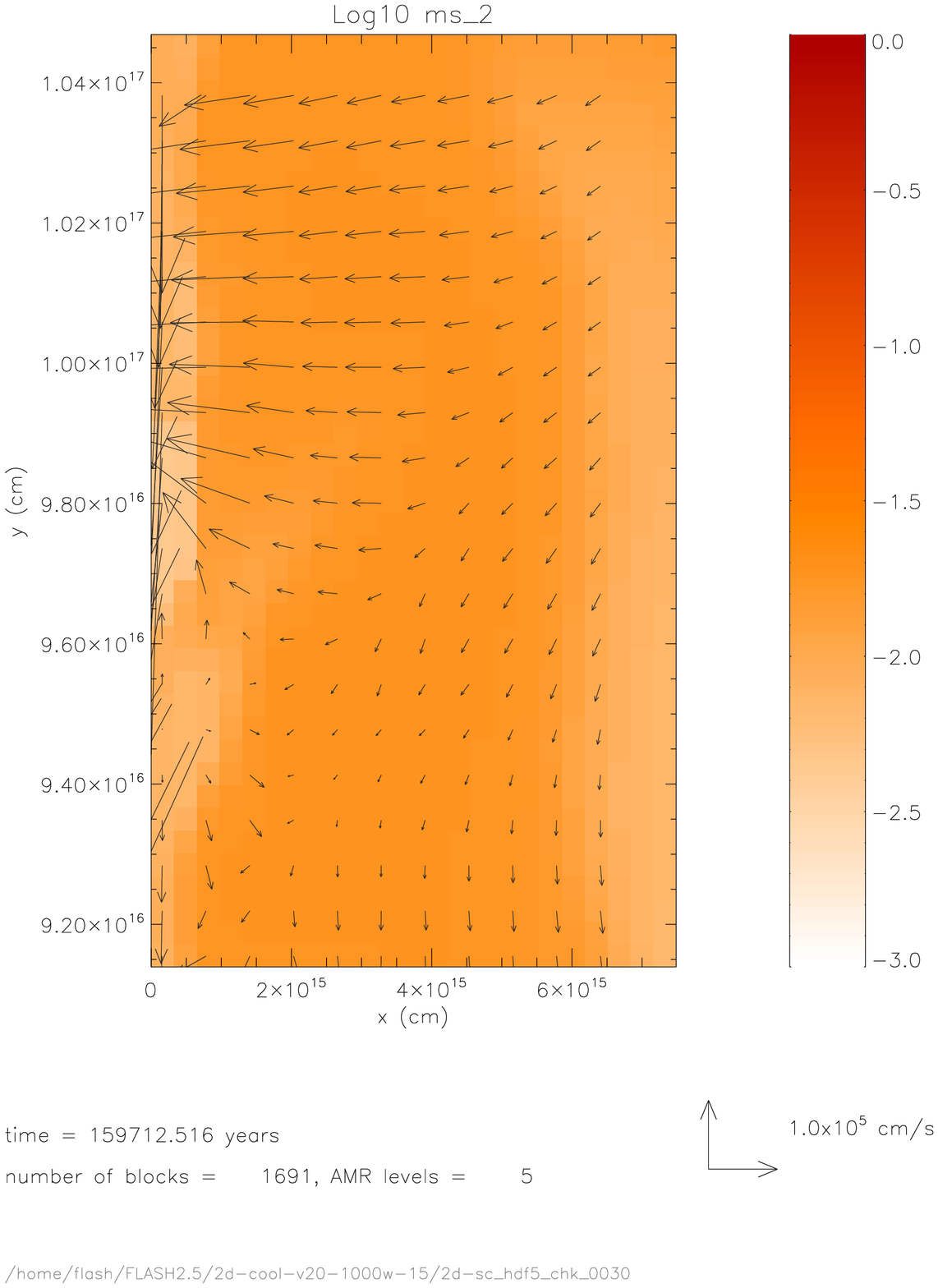}
\vspace{+0.25in}
\caption{Same as Figure 3 except now the color field is plotted
along with the velocity vectors, showing that while the growing
protostar contains some color, the infalling regions contain
a higher color density, i.e., more a higher density of SLRI from
the shock wave.}
\end{figure}

\clearpage

\end{document}